# Free Energy Computation by Monte Carlo Integration

Matthew Clark*
Elsevier R&D Solutions 1600 John F. Kennedy Blvd, Suite 1800 Philadelphia, PA 19103
m.clark@elsevier.com
Jeffrey S. Wiseman
Edison Biotechnology Institute, Ohio University, Athens OH, 45701
Wisemaj2@ohio.edu

## Abstract

The principles behind the computation of protein-ligand binding free energies by Monte Carlo integration are described in detail. The simulation provides gas-phase binding free energies that can be converted to aqueous energies by solvation corrections. The direct integration simulation has several characteristics beneficial to free-energy calculations. One is that the number of parameters that must be set for the simulation is small and can be determined objectively, making the outcome more deterministic, with respect to choice of input conditions, as compared to perturbation methods. Second, the simulation is free from assumptions about the starting pose or nature of the binding site. A final benefit is that binding free energies are a direct outcome of the simulation, and little processing is required to determine them.

The well-studied T4 lysozyme experimental free energy data and crystal structures were used to evaluate the method.

## 1.  Introduction

Free energy computations promise the ability to compute or at least rank binding energies of ligands to a protein structure. Recent work has demonstrated that free energy perturbation computations with modern force fields can achieve high accuracy over a range of protein and ligand types.[1]  In this work we investigate a method to compute free energies that does not use perturbation, or thermodynamic integration. Instead, it uses a direct integration by of the binding free energy by random attempts at physical insertion of the ligand into protein binding locations.

The free energy is computed by integrating the entropy by using the ratio of accepted poses to attempts, combined with the enthalpy computed using a molecular mechanics force-field. The method is similar to that used in systematic sampling.[2] However, random sampling of poses is used, enabling the method to be used on conformationally flexible ligands without introducing assumptions for assembling smaller fragments.

The L99A mutant of T4 lysozyme was created to study protein-ligand binding in an artificial and fully-enclosed hydrophobic pocket.[3] Morton and Matthews reported calorimetrically measured binding free energies and enthalpies for a small series of ligands that have been used by many researchers as a reference point to test calculations of binding affinity.[4,5] The binding to T4 has been studied using dynamics methods [6,7], the blurring method[8], docking[9], replica exchange methods[10], perturbation[11], BEDAM sampling[12], Grand Canonical free energy computations[13,2,14,15], distributed replica sampling[16] and Hamiltonian replica exchange[17,18] among others.

The free energy methods in most common use generally use some form of perturbation whereby a ligand either 'fades in' or 'fades out' of interaction with the binding site, and the energy changes are sampled during dynamics simulations or Monte Carlo moves.[19]

An important limitation of the free energy perturbation methods is that they carry out the computations for a pre-selected ligand pose in a pre-selected binding site of the host. While the dynamics-based sampling regimen allows the ligand to move about within the energy barriers defined by the simulation temperature, the ligand cannot easily cross these barriers to explore different poses and conformations. A significant operational issue results from the tendency of the ligand to leave the vicinity of the host as the ligand potential is decreased in the integration cycle. The use of complex constraints and associated methods to remove the thermodynamic effects of the constraints has been developed to attempt to address this problem.[7] In dynamics methods highly symmetric molecules like benzene require symmetry adjustment factors to correctly estimate entropy.[7]

An advantage of the method described herein is that the simulation is relatively free from assumptions about the starting pose or nature of binding site. The simulation described herein samples and computes the free energy ligands with few assumptions. Any symmetry factors are incorporated into the random sampling. This makes the method well adapted to locating novel binding modes. In the simple cases demonstrated in T4 lysozyme the crystallographic binding site is the lowest energy binding location when the entire protein surface and interior is sampled.

## 2. Methods

### 2.1 Free Energy Computation

The Helmholtz free energy is computed over all sampled pose states from the bound state Z, and the bath unbound states $Z_0$, as shown in Equations 1 and 2.[20,21]

$$\textbf{1} \qquad Z = \sum_i e^{-E_i/kT}$$

$$\textbf{2} \qquad \Delta G = -kT \ln(\frac{Z}{Z_0})$$

Since changes in volume and pressure are negligible, the Gibbs free energy of binding is approximately equal to the Helmholtz free energy of Equation 2. In this work the reference state is defined to have energy of zero, so that the value of $Z_0$ is unity for each pose. In this simulation the poses are randomly sampled for acceptance one a time so that there is no interaction among the ligands.

### 2.2 Implementation

The "Iron" canonical sampling software was written in Java™[22] to carry out the computations. The large quantity random numbers required for sampling were generated using the Mersenne Twister algorithm[23]. The ligand – host interaction energies were computed using an implementation of the AMBER molecular mechanics force field and specifically the AMBER parm99 parameter set.[24] The ligand-protein interaction energy is computed using Equation 3 which consists of the AMBER force field non-boded term summed over atom pairs, augmented by the CHARMM angularly-dependent hydrogen bond term to reduce the dependence of the energy on the electrostatic term.[25]  A non-bonded distance cutoff of 12 Å was used with a distance-dependent dielectric with a constant of 10.

$$\textbf{3} \qquad E = \sum_{ij} \left( \frac{A}{r_{ij}^{12}} - \frac{B}{r_{ij}^{6}} \right) + \sum \frac{kq_i q_j}{r_{ij}^2} + \sum_{\substack{ij\ (hbond \\ pairs)}} D_{ij} \{ -6r_{ij}^{10} + 5r_{ij}^{12} \} cos^4\ \theta_{DHA}$$

The energy of Equation 3 is used to accept or reject sampled poses using the Metropolis Monte Carlo sampling process.[26]

A grid was used to define the ligand sampling region and define sub-regions or cells that are filled by the protein and cells that are available for sampling. This grid was a simple paralleliped defined by a center and the extent in the x, y, and z dimensions and separated into cells of 0.5Å per side. If the van der Waals radius of any atom of the protein structure fell within a cell, the cell was marked as unavailable for sampling. The geometric center of a ligand pose is required to fall within an available cell in the sampling region. The sampling regions and volumes used for the simulations are given in Table 1 and Table 2.

The sampling was carried out in a straightforward manner. A Cartesian position was randomly chosen in the sampling region and the geometric center of the initial ligand conformation translated to that position. Next a random rotation in all 3 Euler angles was made to the ligand and then a new conformation created by randomly changing the rotations of the single bonds by incrementing each by a random angle between 0° and 360°, followed by a collision check among the ligand atoms. The conformation randomization was repeated until an acceptable conformation was found if ligand atoms collided.

The ligand-protein energy was computed using Equation 3, and then the probability of acceptance computed using Equation 4, then a random number between 0 and 1 was generated.

$$\textbf{4} \qquad \boldsymbol{P_{accept} = e^{-\frac{\Delta E}{kT}}}$$

If the acceptance probability was greater than the random number, the pose was accepted into the thermodynamic ensemble. If the pose is accepted, a counter for the total number of accepted poses is incremented for both the entire sampling region, and for the particular cell that contained the insertion point.

The process was carried out repeatedly. The entropy contribution of limiting the rotation and conformational freedom of the pose in the sampling region is computed using the formula of Equation 5. This equation can be used for the entire sampling region, and to track the entropy associated with each individual cell.

$$\textbf{5} \qquad \text{Entropy}_{rot-conf} = \boldsymbol{kT\,ln}\left(\frac{\sum \text{accepted insertions}}{\sum \textbf{attempted insertions}}\right)$$

The translational entropy is computed by comparing the volume of the sampling region to the average volume per molecule of a 1M gas at the simulation temperature, 1661 $Å^3$. The translational entropy is computed using Equation 6.

$$\mathbf{6} \qquad \boldsymbol{Entropy_{trans}} = \boldsymbol{kT}\ \boldsymbol{ln}\left(\frac{\boldsymbol{integration\ volume\ (\text{Å}^3)}}{\boldsymbol{1661\ \text{Å}^3}}\right)$$

The entropy contribution of Equation 7 is computed both for the entire sampling region, and for each individual sampling cell. For example, if the sampling region available volume is 214 Å$^3$ the translation entropy of limiting the translation volume from 1661 to 214 Å$^3$ is 1.2 kcal/mol. For a sampling cell 0.25Å per side the translational entropy of being limited to that cell is 6.9 kcal/mol. The free energy integrated over the binding site is thus given by Equation 7. With this formula the free energy can be computed for any grid cell by tracking the number of attempts and accepted insertions into that particular cell or for the binding site as a whole by counting all insertions.

$$\mathbf{7} \qquad \boldsymbol{\varDelta G}_{\text{binding}} = \frac{\sum \varDelta E^* e^{-\varDelta E/\text{kT}}}{\sum e^{-\varDelta E/\text{kT}}} - \text{Entropy}_{\text{rot-conf}} - \text{Entropy}_{\text{trans}}$$

In Equation 7 the first term is the Boltzmann-weighted enthalpy over all of the accepted poses, and the second and third terms are the entropy as computed in Equations 6 and 7. In order to help find the lowest energy poses an accepted pose was randomly selected with probability 0.25 per step for translation/rotation with a Gaussian distribution from the existing pose to attempt to find lower energy poses. These moves were not used to update the entropy counts of accepted poses. The free energies were computed two ways. First, for the volume of the cell containing the lowest free-energy pose with the entropy being computed by the counts of accepted/rejected poses for attempts just in that cell. The volume used for Equation 6 was the volume of a single cell. Second, the free energies were also computed for the entire sampling region, with the enthalpy being the Boltzmann-weighted value over all poses, and the entropy computed by considering all insertion attempt accepts/rejections for the sampling volume. The volume used for Equation 6 was the volume of the available sampling region.

The calculations were carried out on two slightly different T4 crystal structures, based on 4W56A, and 186L to assess the impact of different binding site choices. These were chosen because the binding sites were pre-formed to accommodate large ligands in each series, *sec*-butyl benzene, and *n*-butyl benzene respectively. Hydrogens were added and the structures equilibrated using Pymol.[27]

A representation of steps in the thermodynamic cycle of protein-ligand binding is shown in

Figure 1. The computation results in the value for $\Delta G_1$. The experimentally measured value corresponds to -$\Delta G_3$. The difference between the two is $\Delta G_2$ - ($\Delta G_4$ + $\Delta G_5$). As in previous work for this system we made the approximation that $\Delta G_2$ = $\Delta G_4$, which was assumed reasonable for the hydrophobic internal pocket of T4 Lysozyme. However previous work demonstrates that this approximation leads to computed binding free energies about 4 kcal/mol too low.[14]

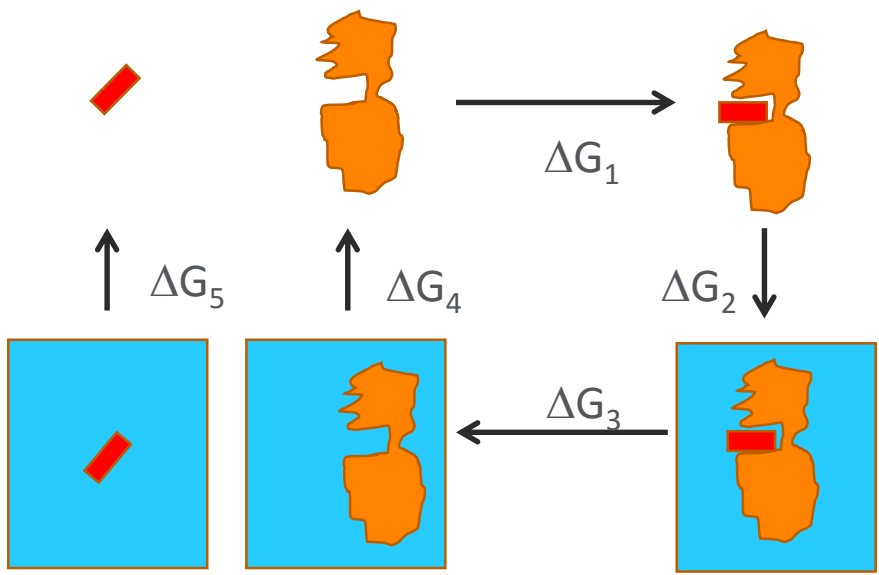

*Figure 1. Thermodynamic Cycle of Protein-Ligand Binding*

The balance of the correction to the observed free energy is the vapor-solvation free energy change of the ligand, $\Delta G_5$. For this work we used the vapor-solvation free energy values computed by the GB/SA method as used in previous publications.

In this work each ligand was sampled with $10^{10}$ insertion attempts into sampling region to insure adequate entropy sampling, and each protein-ligand simulation was carried out 7 separate times to assess the reproducibility of the results. In addition, the process was carried out with two variations of the protein structure to assess the impact of slight changes in the binding site geometry.

In a separate experiment, benzene was sampled over the entire protein volume, a sampling volume of 15,461.5 $\text{Å}^3$ to compare results of the free-energy computation of the entire protein vs sampling only a selected binding site.

*Table 1. Binding Site Sampling Region 186L*

|  | **(values in Å)** | | |
|---|---|---|---|
|  | **x** | **y** | **z** |
| Size | 9.0 | 9.0 | 8.5 |
| Minimum | 23.5 | 2.0 | -0.5 |
| Maximum | 32.5 | 11.0 | 8.0 |
| center | 28.0 | 6.50 | 3.75 |

|  | **Volume** | **Cells** |
|---|---|---|
| Total volume | 688.5 Å$^3$ | 5,508 cells |
| Containing protein atoms | 548.25 Å$^3$ | 4,386 cells |
| Free for Sampling | 140.25 Å$^3$ | 1,122 cells |

*Table 2. Entire Protein Sampling Region 186L*

|  | **(values in Å)** | | |
|---|---|---|---|
|  | **x** | **y** | **z** |
| Size | 49.5 | 52.5 | 61.5 |
| Minimum | 7.5 | -12.0 | -20.0 |
| Maximum | 57.0 | 40.5 | 41.5 |
| center | 32.25 | 14.25 | 10.75 |

|  | **Volume** | **Cells** |
|---|---|---|
| Total volume | 15,9823.125 Å$^3$ | 1,278,585 cells |
| Containing protein atoms | 14,4361.625 Å$^3$ | 1,154,893 cells |
| Free for Sampling | 15,461.5 Å$^3$ | 123,692 cells |

## 2.3   Limitations

While the method is thermodynamically rigorous, the simulation as implemented for this work has limitations that represent areas that could be improved.  The first set of limitations is around the molecular mechanics force field in general, and the AMBER force field used for this simulation. The limitations of molecular mechanics are well known, and recent work has shown that high quality force fields can provide excellent results.[1]  In this work no implicit or explicit solvent model was used, which may suffice for the non-polar ligands and cavity but may be required for good results for other interactions that rely more heavily on hydrogen bonding. While conformations were sampled, the conformational energy beyond simple hard-sphere atom collisions was ignored. This might be a limitation, although some authors find little relationship between computed conformational energy and binding energy.[28,29]

The second set of limitations centers on the assumption of rigid protein, and rigid rotor ligands. In this study the protein structure was fixed, although the ability of the T4 pocket to move to accommodate ligands has been well studied both crystallographically and computationally.[4],[30],[10] The flexibility of VAL111 in T4 lysozyme has been identified as key to binding of some ligands, particularly *n*-butyl benzene and *iso*-butyl benzene. However this study used the crystallographic sidechain conformation.[31] While conformational flexibility was allowed for the ligands, their bond lengths and angles were rigid. Allowing the protein structure to accommodate the ligands through periodic Monte-Carlo searching may improve the results.

## 3. Results

The computed free energies, enthalpies, and entropies for the series of compound measured calorimetrically, using the 186L structure, are shown in Table 3. The values computed with the 4W56 structure are shown in Table 4. The errors for the three thermodynamic values for the two models are summarized in Table 5. The rms prediction error for the linear fitted data to experiment is low, about 0.5 kcal/mol, however it is unclear whether this fit (slope, intercept) is applicable to other systems. The absolute errors of prediction are offset by about 10 kcal/mol. The set of poses of each in the 186L model are shown in Figure 2. The poses sampled over the entire protein are shown in Figure 3.

*Table 3. Simulation data from 186L Structure. Energy values for the lowest free-energy cell in kcal/mol*

| ligand | ΔG | | | ΔH | | | -TΔS | | |
|---|---|---|---|---|---|---|---|---|---|
| 1,2-xylene | -7.84 | ± | 0.073 | -20.73 | ± | 0.193 | -12.89 | ± | 0.233 |
| 1,3-xylene | -8.47 | ± | 0.055 | -21.17 | ± | 0.012 | -12.70 | ± | 0.052 |
| 1,4-xylene | -7.71 | ± | 0.141 | -20.69 | ± | 0.075 | -12.98 | ± | 0.122 |
| 2-ethyltoluene | -8.81 | ± | 0.222 | -23.37 | ± | 0.277 | -14.56 | ± | 0.164 |
| 3-ethyltoluene | -8.28 | ± | 0.220 | -22.93 | ± | 0.195 | -14.65 | ± | 0.223 |
| 4-ethyltoluene | -9.14 | ± | 0.202 | -23.34 | ± | 0.146 | -14.20 | ± | 0.183 |
| benzene | -7.99 | ± | 0.293 | -16.93 | ± | 0.309 | -8.94 | ± | 0.081 |
| benzofuran | -11.99 | ± | 0.134 | -22.52 | ± | 0.138 | -10.53 | ± | 0.091 |
| ethylbenzene | -10.96 | ± | 0.038 | -22.50 | ± | 0.038 | -11.54 | ± | 0.026 |
| indene | -9.85 | ± | 0.118 | -21.21 | ± | 0.120 | -11.36 | ± | 0.024 |
| indole | -11.68 | ± | 0.009 | -22.37 | ± | 0.012 | -10.69 | ± | 0.018 |
| *iso*-butylbenzene† | - | | | - | | | - | | |
| naphthalene | -8.99 | ± | 0.045 | -21.98 | ± | 0.042 | -12.99 | ± | 0.042 |
| *n*-butylbenzene* | -11.70 | | | -26.94 | | | -15.25 | | |
| *n*-propylbenzene | -12.33 | ± | 0.251 | -26.14 | ± | 0.218 | -13.81 | ± | 0.179 |
| pyridine | -8.63 | ± | 0.522 | -17.35 | ± | 0.463 | -8.72 | ± | 0.155 |
| *t*-butylbenzene† | - | | | - | | | - | | |
| thianaphthene | -11.67 | ± | 0.232 | -22.84 | ± | 0.375 | -11.16 | ± | 0.349 |
| toluene | -10.11 | ± | 0.129 | -19.79 | ± | 0.366 | -9.68 | ± | 0.241 |

*in all 7 runs only one attempt was accepted

† no insertion attempts were accepted in any run

*Table 4. Simulation Data from 4W56 Structure. Energy values for the lowest free-energy cell in kcal/mol.*

| ligand | ΔG | | | ΔH | | | -TΔS | | |
|---|---|---|---|---|---|---|---|---|---|
| 1,2-xylene | -10.44 | ± | 0.012 | -21.52 | ± | 0.009 | -11.08 | ± | 0.018 |
| 1,3-xylene | -10.21 | ± | 0.442 | -21.02 | ± | 0.685 | -10.81 | ± | 0.260 |
| 1,4-xylene | -9.21 | ± | 0.039 | -20.56 | ± | 0.190 | -11.36 | ± | 0.154 |
| 2-ethyltoluene | -11.22 | ± | 0.109 | -24.18 | ± | 0.150 | -12.96 | ± | 0.104 |
| 3-ethyltoluene | -11.84 | ± | 0.093 | -23.95 | ± | 0.284 | -12.11 | ± | 0.238 |
| 4-ethyltoluene | -11.47 | ± | 0.097 | -23.55 | ± | 0.126 | -12.09 | ± | 0.105 |
| benzene | -7.96 | ± | 0.814 | -17.01 | ± | 0.978 | -9.05 | ± | 0.280 |
| benzofuran | -12.19 | ± | 0.555 | -21.92 | ± | 0.676 | -9.74 | ± | 0.222 |
| ethylbenzene | -11.20 | ± | 0.122 | -21.95 | ± | 0.115 | -10.74 | ± | 0.013 |
| indene | -11.87 | ± | 0.097 | -22.12 | ± | 0.098 | -10.25 | ± | 0.089 |
| indole | -12.28 | ± | 0.408 | -22.27 | ± | 0.561 | -9.99 | ± | 0.180 |
| *iso*-butylbenzene | -11.53 | ± | 0.410 | -26.65 | ± | 0.396 | -15.12 | ± | 0.081 |
| naphthalene | -13.00 | ± | 0.197 | -23.50 | ± | 0.196 | -10.50 | ± | 0.013 |
| *n*-butylbenzene | -14.08 | ± | 0.121 | -28.25 | ± | 0.016 | -14.17 | ± | 0.121 |
| *n*-propylbenzene | -12.88 | ± | 0.167 | -24.79 | ± | 0.303 | -11.91 | ± | 0.170 |
| pyridine | -9.14 | ± | 0.720 | -17.56 | ± | 0.791 | -8.43 | ± | 0.197 |
| *t*-butylbenzene† | - | | - | - | | - | - | | - |
| thianaphthene | -13.14 | ± | 0.537 | -23.03 | ± | 0.626 | -9.89 | ± | 0.110 |
| toluene | -9.54 | ± | 0.226 | -18.94 | ± | 0.244 | -9.40 | ± | 0.107 |

† no insertion attempts were accepted in any run

*Table 5. Errors in binding energy for the two models. Energy values for the lowest free-energy cell; all data in kcal/mol. Experimental values from reference 4.*

| | **186L** | | | **4w56** | | | **Experiment** | | |
| | ΔG error | ΔH error | ΔS error | ΔG error | ΔH error | ΔS error | ΔG | ΔH | ΔS |
|---|---|---|---|---|---|---|---|---|---|
| 1,2-xylene | -3.87 | -12.72 | -8.85 | -5.84 | -13.07 | -7.23 | -4.60 | -8.45 | -3.85 |
| 1,3-xylene | -3.09 | -14.69 | -11.60 | -5.46 | -14.98 | -9.52 | -4.75 | -6.04 | -1.29 |
| 1,4-xylene | -3.04 | -13.72 | -10.68 | -4.53 | -13.59 | -9.05 | -4.67 | -6.97 | -2.30 |
| 2-ethyltoluene | -4.25 | -15.66 | -11.41 | -6.66 | -16.47 | -9.81 | -4.56 | -7.71 | -3.15 |
| 3-ethyltoluene | -3.16 | -15.22 | -12.06 | -6.72 | -16.24 | -9.52 | -5.12 | -7.71 | -2.59 |
| 4-ethyltoluene | -3.72 | -14.90 | -11.18 | -6.05 | -15.11 | -9.07 | -5.42 | -8.44 | -3.02 |
| benzene | -2.80 | -10.61 | -7.81 | -2.64 | -10.52 | -7.88 | -5.19 | -6.32 | -1.13 |
| benzofuran | -6.53 | -14.48 | -7.95 | -6.62 | -13.86 | -7.24 | -5.46 | -8.04 | -2.58 |
| ethylbenzene | -5.20 | -15.74 | -10.54 | -5.44 | -15.19 | -9.74 | -5.76 | -6.76 | -1.00 |
| indene | -4.72 | -12.90 | -8.18 | -6.74 | -13.81 | -7.07 | -5.13 | -8.31 | -3.18 |
| indole | -6.79 | -11.14 | -4.35 | -7.39 | -11.04 | -3.65 | -4.89 | -11.23 | -6.34 |
| n-butylbenzene | -5.00 | -18.88 | -13.88 | -7.38 | -20.19 | -12.81 | -6.70 | -8.06 | -1.36 |
| n-propylbenzene | -5.78 | -16.17 | -10.39 | -6.33 | -14.82 | -8.49 | -6.55 | -9.97 | -3.42 |
| thianaphthene | -5.96 | -15.81 | -9.84 | -7.43 | -16.00 | -8.57 | -5.71 | -7.03 | -1.32 |
| toluene | -4.59 | -13.26 | -8.67 | -3.96 | -12.35 | -8.38 | -5.52 | -6.53 | -1.01 |
| | | | | | | | | | |
| *std error of fit* | 0.44 | | | 0.55 | | | | | |
| *average absolute error* | -9.90 | | | -11.28 | | | | | |

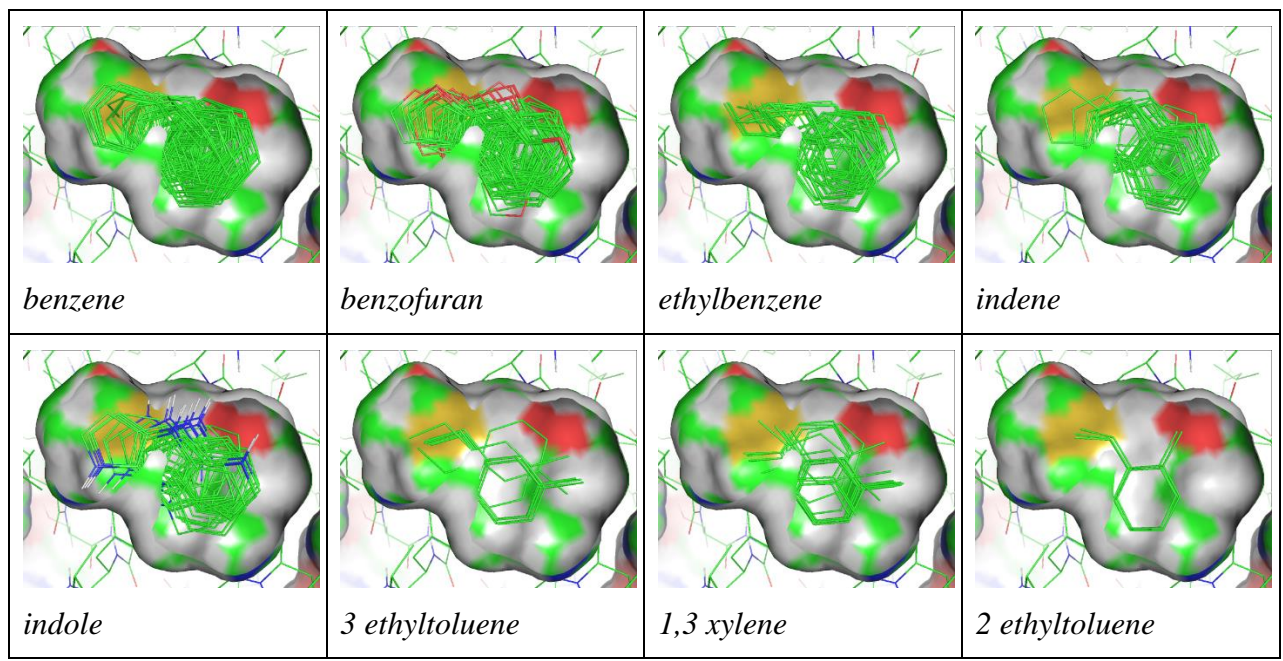

| | |
|---|---|
| *benzene* | *benzofuran* |

*ethylbenzene*

*indene*

*indole*

*3 ethyltoluene*

*1,3 xylene*

*2 ethyltoluene*

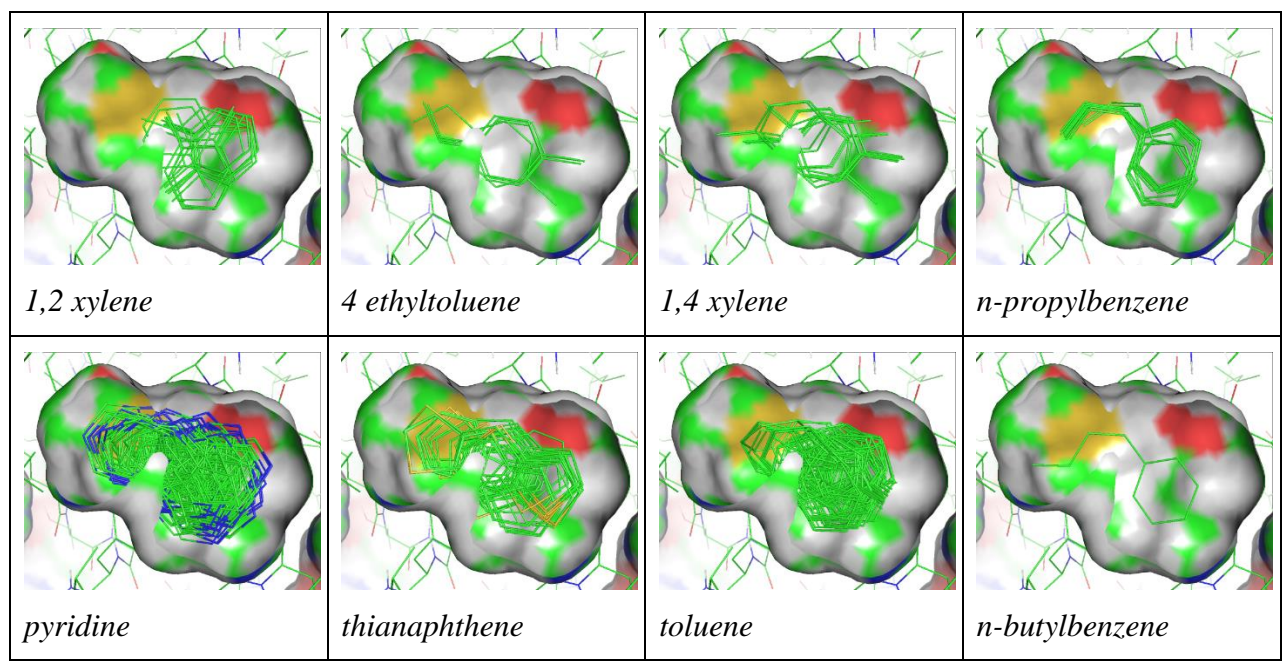

*Figure 2. Ligand poses in the binding site*

*Table 6. Comparison of Energies from cell, binding site, and full protein sampling in 186L.*

| *Integration volume* | Volume (Å$^3$) | (kcal/mol) ΔG | ΔH | TΔS | Count of Poses |
|---|---|---|---|---|---|
| benzene | | | | | |
|    full protein | 15461.5 | -15.2 | -15.7 | -0.4 | 100,186 |
|    binding site | 140.25 | -9.5 | -15.8 | -6.4 | 119 |
|    lowest energy cell | 0.125 | -8.4 | -16.2 | -7.8 | 1 |
| ethylbenzene | | | | | |
|    full protein | 15461.5 | -18.9 | -20.4 | -1.5 | 90,721 |
|    binding site | 140.25 | -10.4 | -22.2 | -11.9 | 53 |
|    lowest energy cell | 0.125 | -11.1 | -22.5 | -11.4 | 1 |

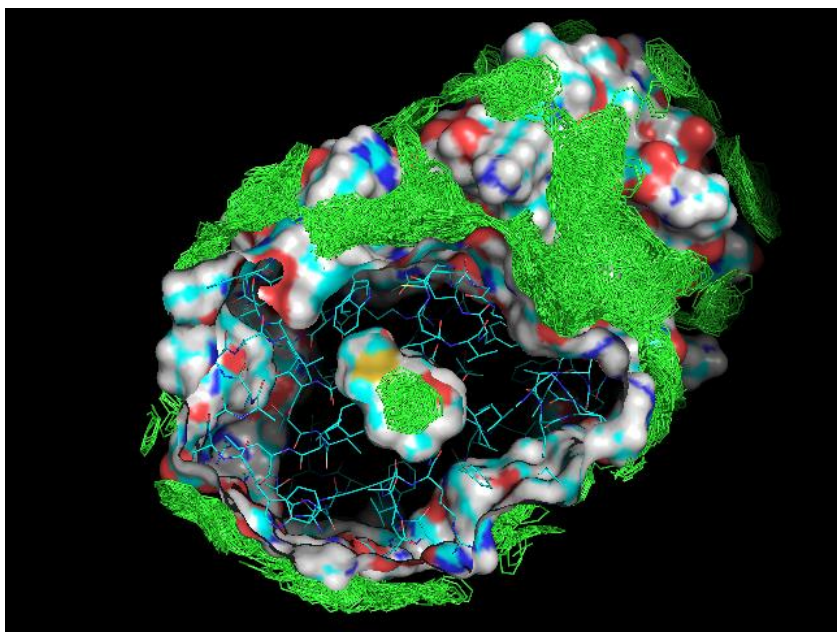

*Figure 3. Benzene sampling over complete protein.*

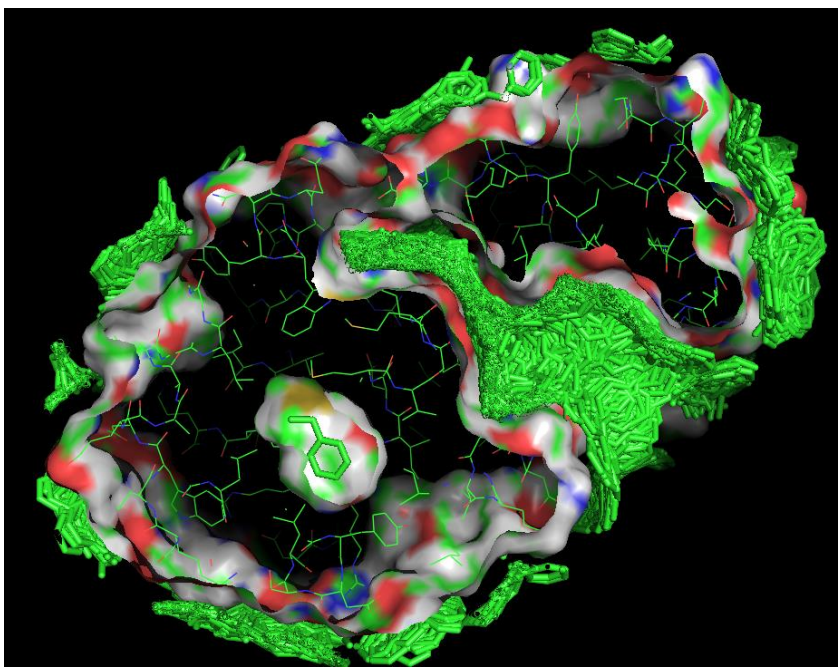

*Figure 4. Ethylbenzene sampling over complete protein.*

# 4. Discussion

## 4.1 Effect of free-energy integration volume

Table 6 shows the differences in integrated free energy for three different integration volumes.

For each volume the system enthalpy is computed using a Boltzmann-weighted average over all of the poses in the volume. As a result, the larger volumes have slightly more positive enthalpy due to the contributions of higher-energy poses. In addition, the larger volumes result in more positive entropies because a larger fraction of poses have low energies and are accepted around the surface of the protein.

The difference between using entropy values integrated over a single cell or the entire sampling volume is shown in Figure 5. They are highly correlated. The conclusion is that for this binding site the free energy integrated over the cell containing the lowest free-energy pose is similar to the energy integrated over the entire binding site. If a binding site had multiple low energy poses this might not be the case. In that situation the value integrated over the entire binding site might be more representative of the experimental binding.

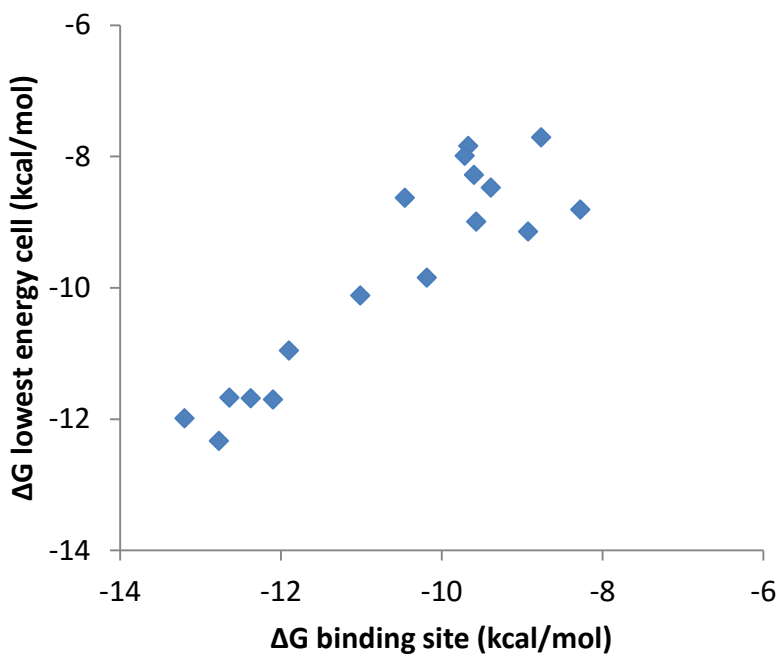

*Figure 5. Free energy comparison computed using data integrated over binding site and lowest energy cell.*

## 4.2   Effect of Protein Model

The exact model used for the protein had a demonstrable effect on result. The slightly larger cavity model generated by starting with the 4W56 (146.5 Å$^3$) structure did not provide as good result as the smaller cavity of the 186L model (140.25 Å$^3$). A limitation of this method is that the cavity must be large enough for the ligand to fit to be sampled efficiently. Closely fitting ligands are not likely to be accepted and therefore have high entropy. In this study the slightly larger pocket resulted in similar interaction energies, as shown in Figure 10, but generally higher entropy values as shown in Figure 9. The ligands with the largest differences are *n*-butyl benzene, which was accepted only once in $7x10^{10}$ steps in the 186L model, and *iso*-butyl benzene which was accepted in the 4W56 model but not the 186L model. The overall entropies limited to a single cell are much higher than the measured values, although the range is similar. There is no other correlation between the predicted and actual entropy values.

## 4.3   Reproducibility

Publications reporting the reproducibility of free energy calculations for specific protein-ligand pairs are rare. Recently the Merz group has reported calculated uncertainties for T4 ligands using the "Blurring" method.[8,32] Since Monte Carlo simulations involve random sampling one would expect some variation in the results of complex calculations. In order to investigate the magnitude of the effects with the sampling protocol and software, each protein-ligand simulation was carried out seven times.

Although each simulation was run for $10^{10}$ steps, not every simulation found the same lowest energy pose. The average deviation in free energy or enthalpy was 2%. Pyridine and benzene had the highest variation at 6 and 4% respectively. This may be due to the freedom to move about in the binding site; there are more low energy poses near the lowest energy pose available.

The relatively low variation in energies is reassuring that the method, and the sampling protocol, are providing as reliable a result as the force field and structural models allow.

## 4.4   Sampling region size

Experimentally measured free energy is not specifically for a 'binding site'; it is the overall binding where the lowest energy pose (normally in a known binding site) is expected to more populated than other poses with a frequency exponentially related to the binding energy. Because of that one would hope that the energy function could be used to sample the entire protein area and correctly identify the binding pose.

In fact, for this very simple case this is the outcome. Sampling over a volume that contains the entire protein locates the same lowest-energy poses as sampling only the known binding site.

The poses for benzene and ethyl benzene are shown in Figure 3 and Figure 4. However, this might not be the case for more polar molecules interacting with charged residues on the protein surface. A comparison of the energies obtained for the binding site and the entire protein for benzene is shown in Figure 3, that for ethyl benzene in Figure 4

Free energy associated with binding site is similar regardless of overall sampling area. Overall lowest free energy location is the same, but sampling not as efficient for whole protein sampling. Entropy assignment is more philosophical; best result is likely entropy of binding region.

### 4.5   Sampling convergence

Sampling is relatively slow even for rigid molecules like benzene. Even for a relatively small sampling region Figure 6. For a more flexible molecule the convergence was paradoxically faster, as seen for propylbenzene shown in Figure 7. As discussed for entropy, this may be because the small benzene ligand has many low-energy poses available while propylbenzene has only a few.

For full protein sampling the results are the same as binding site but more steps are required to sample each cell the same number of times. Because many more poses are accepted on the whole, the system entropy change is less when sampling the entire protein surface.

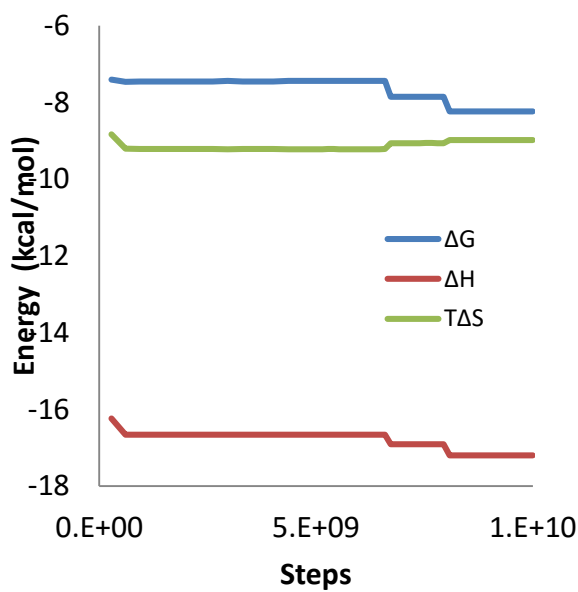

*Figure 6. Energy Convergence of benzene*

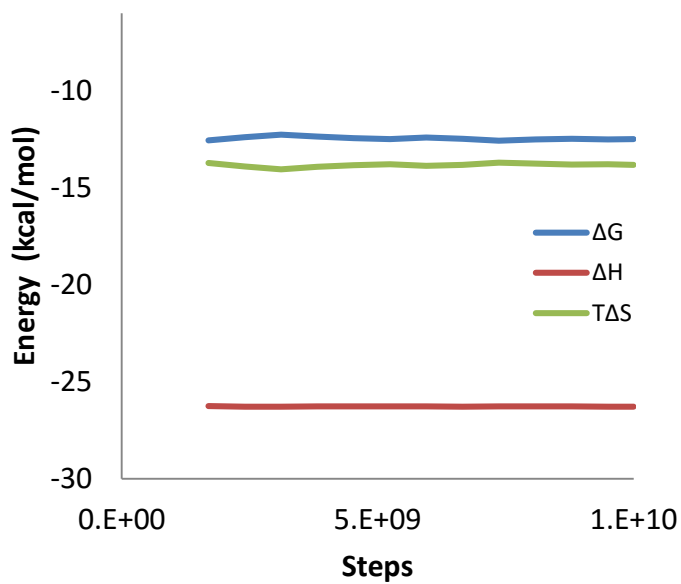

*Figure 7. Energy convergence of propylbenzene*

## 4.6 Comparison to experiment

Binding free energy without solvation correction is shown in Figure 11. The correlation between experiment and calculation is moderately good, but improved with solvation correction. The comparison of experimental and computed binding with solvation correction is shown in Figure 12. The results and correlation are similar to previous work. Predicted vs measured enthalpy is shown in Figure 10.

Computed binding enthalpy is offset by 15 kcal/mol and is the slope 0.88 when plotted against measured values. The solvation energy correction does not bring the scale of computed free energies closer to the experimental values. In fact the computed vs observed free energy plot has a slope of 1.9. Applying the solvation correction improves the linear fit, but increases the slope to 2.7. While there is an overall correlation of the free energies the current method and force field clearly does not have the ability to compute the measured values for binding enthalpy, entropy or free energy.

However, the calculation captures the general relationship between enthalpy and entropy, with the trend that ligands that bind with higher enthalpy generally have higher entropy. This comparison is shown in Figure 8. This nature and origin of this compensation has been widely discussed, and relationship is clearly reproduced by this simulation method.[33,34,35,36]

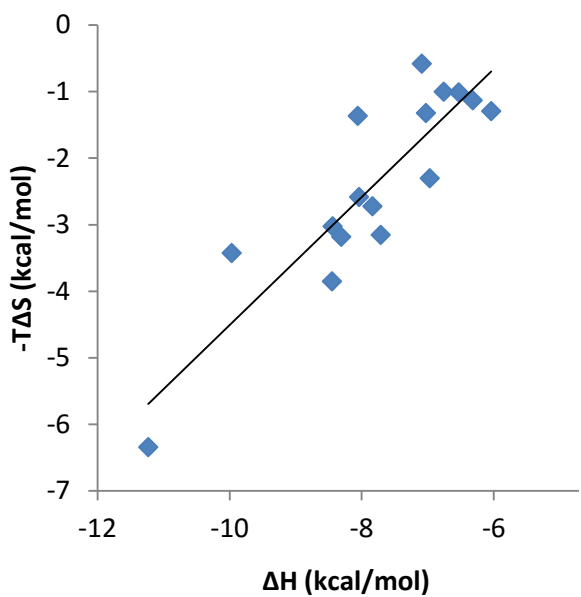 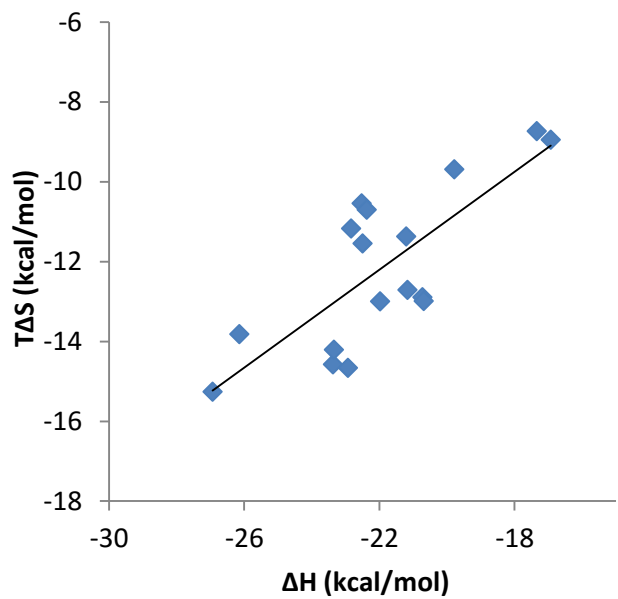

Experimental values                    Computed values

*Figure 8. Enthalpy/entropy compensation, experimental and computed.*

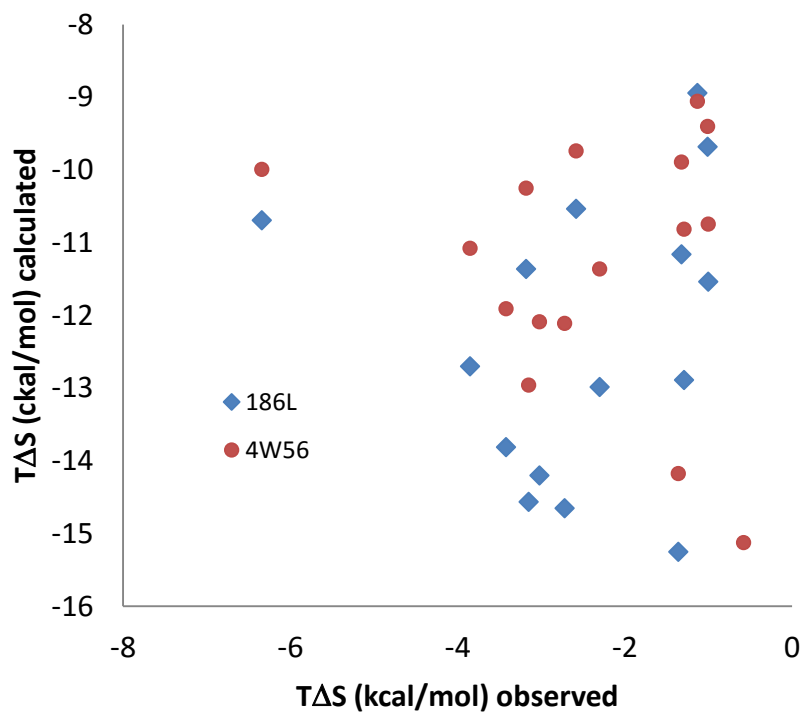

*Figure 9. Computed vs Experimental Entropy*

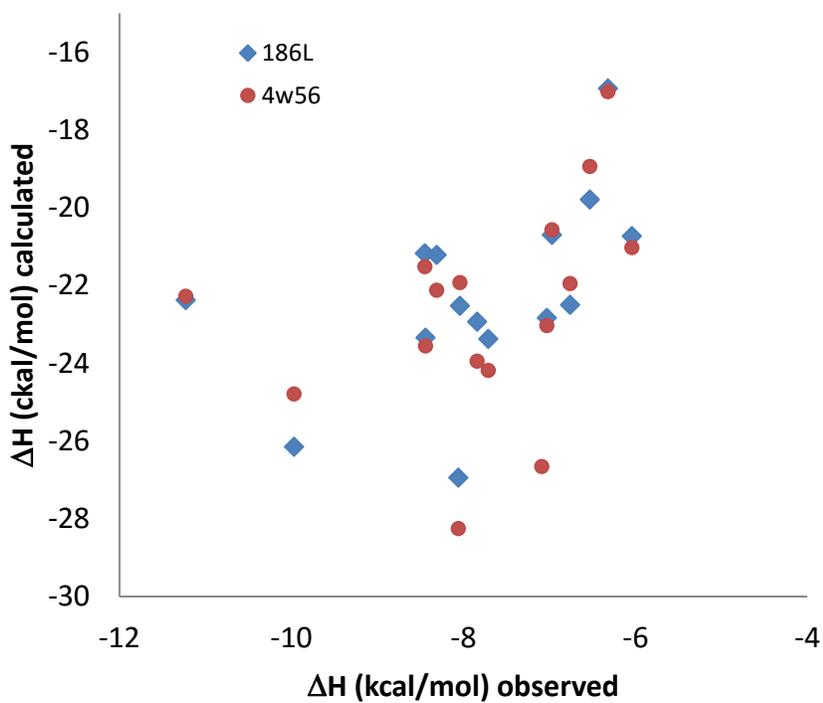

*Figure 10. Computed vs Experimental Enthalpy*

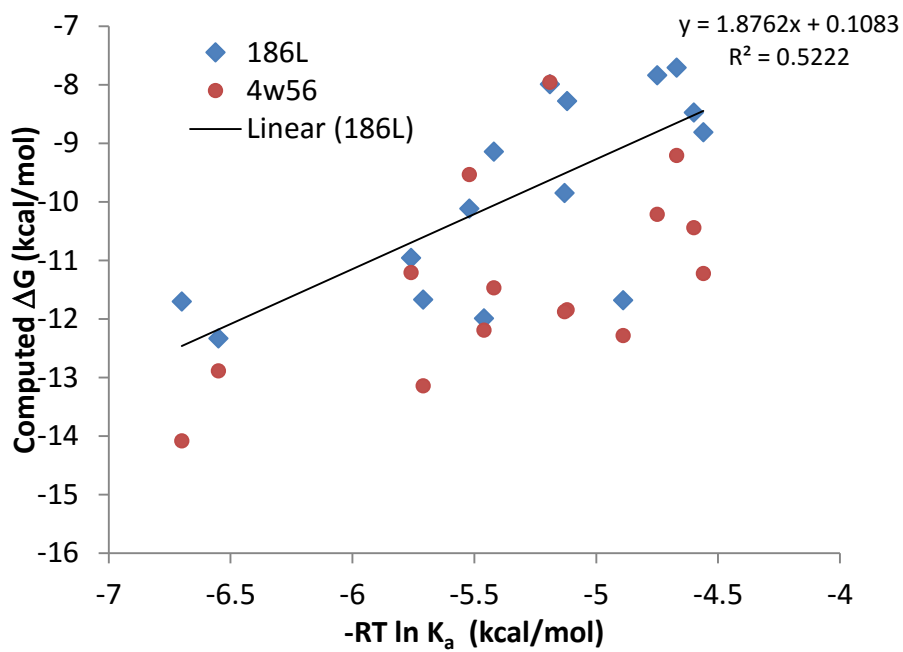

*Figure 11. Predicted vs Actual Binding Free Energy, no solvation correction.*

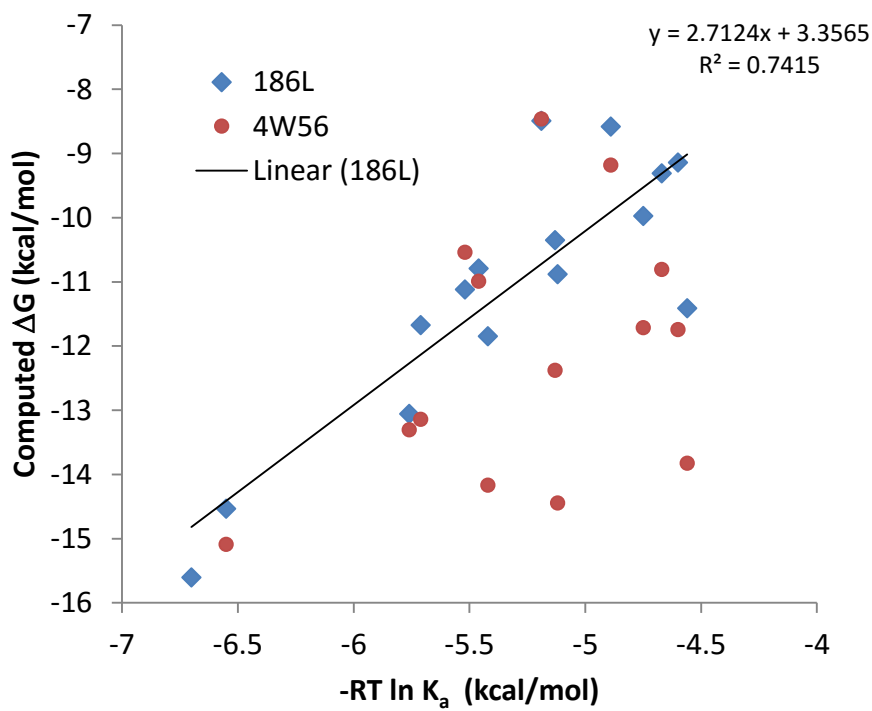

*Figure 12. Predicted vs Actual Binding Free Energy with Solvation Correction*

## 4.7    Comparisons to previous work

In previous work the same protein and ligand set was used to integrate free energies using assembled fragments.[2] In this case the calculated entropies by direct integration are remarkably similar to those computed by the combinations of two and three fragments.  Both methods compute an entropy of 14.2 kcal/mol for 4-ethyl toluene.  For n-butyl benzene the random sampling resulted in an enthalpy of -15.3 kcal/mol with one sample in $7 \times 10^{10}$ steps, while the fragment assembly process resulted in a lower -19.0 kcal/mol. This illustrates a limitation of this method for high-entropy molecules; using Equations 5  and 6 an acceptance every $2 \times 10^{10}$ steps is required to obtain an entropy of -20 kcal/mol.  For enthalpy calculations the values in this work are slightly lower, possibly because methyl groups are allowed to rotate to optimize the energy.

Table 7 compares values reported for benzene by various groups that have used this system to test free-energy calculations. In most cases the solvation-uncorrected binding free energies are close to 9 kcal/mol. Those calculations using solvation models are generally much closer to the absolute measured value.  The main difference between the work of reference 14 and this work is that the assumption that $\Delta G_2 = \Delta G_4$ from Figure 1 was not made; explicit computations were done to compute the two solvation values, which resulted in an energy correction of 4.1 kcal/mol, placing the value much closer to the experimental value.  The binding free energy of water, using the tip4p model is -3.11 kcal/mol.[37] If this is used to offset the computed energy the benzene binding energy is -5.38, which is close to the computed value.  However, that offset does not improve the overall agreement and wider range of computed than experimentally measured values.

*Table 7. Comparison of binding energies of benzene computed by various free-energy methods.*

| Source | Reference | ΔG (kcal/mol) |
|---|---|---|
| *experimental* | 5 | -5.19 |
| this work | | -8.49 |
| Jo, Jiang | [32] | -9.77 |
| Ucisik, Zheng, Faver et. al. | 8 | -8.63 |
| Gallichio | [38] | -4.01 |
| Wang, Chodera | 10 | -4.26 |
| Mobley, Graves, Chodera | [31] | -4.56 |
| Deng, Roux | [39] | -5.96 |
| Clark 2009 | 14 | -9.40 |
| Clark 2005 | 15 | -9.81 |
| Clark 2009 | 2 | -4.20 |
| Malmstrom, Watowich | 9 | -5.20 |

# 5. Conclusions

The Monte Carlo integration method provides a robust way to compute thermodynamic binding free energies with few initial assumptions beyond protein structure and ligand geometry. The main limitations are that the protein model must be able to accept the ligands, and that highly flexible ligands, or those that fit very precisely into the binding pocket may require extended sampling to estimate entropy. The lack of assumptions and concomitant simple calculation set-up makes this method ideal for automated calculations of large numbers of ligands binding to a protein. It may also be a method to locate binding poses that can be used as the starting points for the more common free energy perturbation calculations.

The absolute free-energies suffer without a more advanced solvation energy correction; the method may be improved by the addition of solvation terms. A second area of improvement is to allow the protein to move flexibly in response to the ligand. This has been observed to be significant factor for T4 ligands, and may increase the quality of the results.

# 6. Funding

This research did not receive any specific grant from funding agencies in the public, commercial, or not-for-profit sectors.